\documentclass[final,5p,times,twocolumn,preprint]{elsarticle}

\usepackage[utf8]{inputenc}
\usepackage{amssymb}

\DeclareUnicodeCharacter{03C7}{\ensuremath{\chi}}
\DeclareUnicodeCharacter{03B7}{\ensuremath{\eta}}
\DeclareUnicodeCharacter{03C0}{\ensuremath{\pi}}
\usepackage{amsmath}
\usepackage{overpic}
\usepackage{graphicx}
\usepackage{subfigure}
\usepackage{dcolumn}
\usepackage{bm}
\usepackage{epsfig}
\usepackage{xcolor}
\usepackage{enumerate}
\usepackage{booktabs}
\usepackage{multirow}
\usepackage{braket}
\usepackage{tikz}
\usepackage{comment}
\usepackage{orcidlink}

\hypersetup{colorlinks,
            urlcolor=blue,
            linkcolor=blue,
            anchorcolor=blue,
            citecolor=blue}

\newcommand{\BESIIIorcid}[1]{\href{https://orcid.org/#1}{\hspace*{0.1em}\raisebox{-0.45ex}{\includegraphics[width=1em]{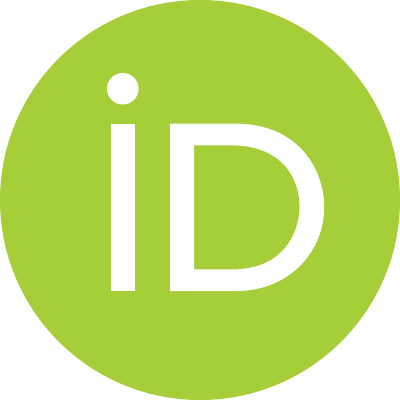}}}} 

\journal{Science Bulletin}

\begin{document}

\begin{frontmatter}

\title{Observation of a Threshold Enhancement in the $\pi^0\pi^0$ Mass Spectrum in $\psi(3686) \rightarrow \pi^{0}\pi^{0}J/\psi$ Decays}

\author{BESIII Collaboration\footnotemark[1]$^{,}$\footnotemark[2]}

\begin{abstract}
Using a data sample of $(2712.4\pm14.4)\times 10^{6}$ $\psi(3686)$ events collected with the BESIII detector, we report a high-precision study of the $\pi^{0}\pi^{0}$ mass spectrum from  $\psi(3686)\rightarrow\pi^{0}\pi^{0}J/\psi$ decays. The enhancement previously observed near the $\pi\pi$ mass threshold in the $\pi^+\pi^-$ channel is confirmed, with a statistical significance exceeding 10$\sigma$. 
A fit with a Breit-Wigner function yields a mass of $M = 274.3\pm 2.1_{\rm stat.}\pm0.6_{\rm syst.}~{\rm MeV}/c^2$ and a width of $\Gamma = 17.1\pm 1.5_{\rm stat.}\pm0.5_{\rm syst.}~{\rm MeV}$, consistent with those from the $\pi^{+}\pi^{-}$ channel.
By including final-state interactions, we analyze the data using both chiral perturbation theory (ChPT) and QCD multipole expansion (QCDME). While the ChPT model describes the higher-mass region above 0.3 GeV/$c^2$ well, it fails to reproduce the significant  threshold enhancement. In contrast, the QCDME model, which incorporates the $S$- and $D$-wave admixture structure of $\psi(3686)$, provides a better description of the full $\pi^{0}\pi^{0}$ mass spectrum. 
\end{abstract}


\end{frontmatter}

\footnotetext[1]{Corresponding author (email: besiii-publications@ihep.ac.cn)}
\footnotetext[2]{Authors are listed at the end of this paper.}

\newcommand{\jpsi}{J/\psi}
\newcommand{\pip}{\pi^+}
\newcommand{\pin}{\pi^-}
\newcommand{\pio}{\pi^0}
\newcommand{\g}{\gamma}
\newcommand{\gev}{GeV/c$^2$}
\newcommand{\mev}{MeV/c$^2$}
\newcommand{\ar}{\rightarrow}
\newcommand{\ks}{K_S^{0}}
\newcommand{\etap}{\eta^\prime}


\section{Introduction}

The study of dipionic transitions between heavy quarkonia, such as $\psi(3686)\rightarrow\pi\pi J/\psi$ decays, provides an important testing ground for non-perturbative Quantum Chromodynamics (QCD) and the dynamics of chiral symmetry breaking. In these transitions, the energy difference between the initial and final charmonium states is released via the emission and subsequent hadronization of soft gluons into a pion pair. This process is particularly sensitive to the low-energy structure of QCD, especially the final-state interactions (FSI) between the emitted pions, which are governed by the same physics that dictates low-energy $\pi\pi$ scattering.

Recent high-precision data from the BESIII Collaboration on the charged mode $\psi(3686)\rightarrow\pi^+\pi^- J/\psi$ decays~\cite{BESIII:2025ozb} revealed a puzzling, narrow enhancement near the $\pi^+\pi^-$ mass threshold with a mass of approximately 283 MeV/$c^2$ and a width of about 17 MeV, which challenges conventional descriptions and is interpreted as an admixture of $S$- and $D$-wave charmonium together with FSI. The distinct dip, approaching zero, in the double-resonance-like structures also verifies the Adler zero condition~\cite{Pham:1975zq, Adler:1964um}, providing direct evidence for the Nambu--Goldstone nature of the pion. This finding makes a dedicated investigation of the neutral mode $\psi(3686)\rightarrow\pi^0\pi^0 J/\psi$ exceptionally compelling.  Since Bose symmetry restricts the $\pi^0\pi^0$ system to even isospin ($I=0,2$, no $I=1$), it probes the $\pi\pi$ final-state interaction more cleanly than the mixed-isospin $\pi^+\pi^-$ channel. Additionally, since the neutral pion mass is slightly less than that of the charged one, any anomalous structure observed in the $\pi^0\pi^0$ spectrum around the $\pi^+\pi^-$ mass threshold would strongly favor a cusp effect rooted in pion-pion rescattering dynamics, as expected in Ref.~\cite{Liu:2012dv}. 

The anomalous structures observed in the $\pi^+\pi^-$ invariant mass spectrum have stimulated several theoretical interpretations, including di-pion correlations~\cite{Wang:2025kcy}, possible signatures of exotic mechanisms, complex contributions from chiral dynamics and improved by charmonium-like tetraquark states~\cite{Chen:2025jip}, as well as $J/\psi\pi$ interactions~\cite{Yan:2026oil}.

With $8.4\times 10^{6}$ $\psi(3686)\rightarrow\pi^{0}\pi^{0}J/\psi$ decays selected from a sample of 2.7 billion $\psi(3686)$ events~\cite{BESIII:2024lks}, we report a precision study of the isoscalar $\pi^{0}\pi^{0}$ system, confirming the resonance-like structure previously observed near the $\pi^+\pi^-$ mass threshold~\cite{BESIII:2025ozb}. 

The BESIII detector records symmetric $e^+e^-$ collisions provided by the BEPCII storage ring~\cite{Yu:2016cof} in the center-of-mass energy range from 1.84 to 4.95~GeV. A detailed description can be found in Refs.~\cite{BESIII:2009fln, Jiao:2020dqs, Zhang:2022bdc, Cao:2020ibk}.

\section{Event selection}

To reconstruct $\psi(3686)\rightarrow\pi^{0}\pi^{0}J/\psi$, the $\pi^0$ is observed in its $\gamma\gamma$ decay mode and the $J/\psi$ meson is detected in
the modes $J/\psi\rightarrow e^+e^-$ and $J/\psi\rightarrow\mu^+\mu^-$. Candidate events must have two oppositely charged tracks identified as leptons and at least four photon candidates.
Charged tracks detected in the multilayer drift chamber (MDC) are required to be within a polar angle ($\theta$) range of $|\cos{\theta}|<0.8$ for leptons, to suppress background events from the $e^+e^-$ annihilation processes directly, where $\theta$ is defined with respect to the $z$-axis, the symmetry axis of the MDC. The distance of closest approach to the interaction point must be less than 10 cm along the $z$-axis, and less than 1 cm in the transverse plane. 
For each charged track, the ratio of energy deposited in the electromagnetic calorimeter (EMC) to the momentum measured in the MDC ($E/p$) is used to distinguish muons from electrons. 
Muon candidates are required to satisfy $E/p < 0.26$, while electron candidates are required to satisfy $E/p > 0.8$.

Photon candidates are reconstructed using clusters of energy deposited in the EMC. The energy deposited in the nearby time-of-flight system is included in EMC measurements to improve the reconstruction efficiency and the energy resolution. Photon candidates are required to have deposited energy larger than 25 MeV in the barrel region ($|\cos \theta|<0.80$) and 50 MeV in the endcap regions ($0.86<|\cos\theta|<0.92$). 
To suppress electronic noise and showers unrelated to the event, the difference between the EMC time and the event start time is required to be within [0, 700]\,ns.

A seven-constraint (7C) kinematic fit, imposing energy-momentum conservation together with mass constraints on the $J/\psi$ and two $\pi^0$ mesons, is performed under the $\psi(3686)\rightarrow\pi^0\pi^0 J/\psi$ hypothesis. Candidate events are required to satisfy  $\chi_{\rm 7C}^2 < 100$. 

\begin{figure}[htbp]
    \centering
    \includegraphics[width=0.42\textwidth]{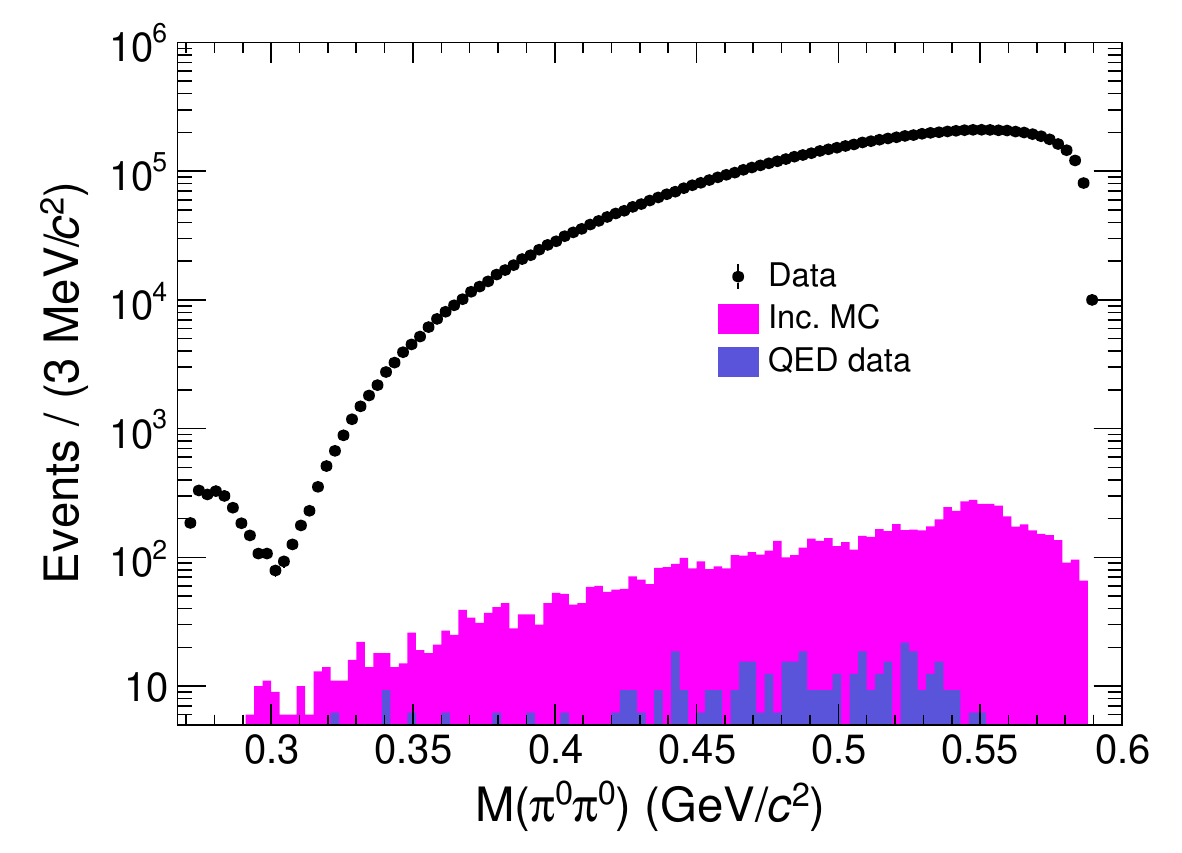}
    \caption{$M(\pi^0\pi^0)$ distribution for $\psi(3686)\rightarrow\pi^0\pi^0J/\psi$ events. The dots with error bars are data. The violet-shaded histogram is the inclusive MC sample. The blue-shaded histogram is the Quantum Electrodynamics (QED) data sample at $\sqrt{s}=3.65~{\rm GeV}$. 
    }
    \label{fig:bkgpipi}
\end{figure}

After applying the above requirements, a sample of $8.4\times10^6$ $\psi(3686)\rightarrow\pi^{0}\pi^{0}J/\psi$ events is selected.
The corresponding $\pi^0\pi^0$ mass spectrum is shown in Fig.~\ref{fig:bkgpipi}, revealing a pronounced enhancement near the $\pi^0\pi^0$ mass threshold. 

\section{Background studies}

\begin{figure}[htbp]
    \centering
    \includegraphics[width=0.42\textwidth]{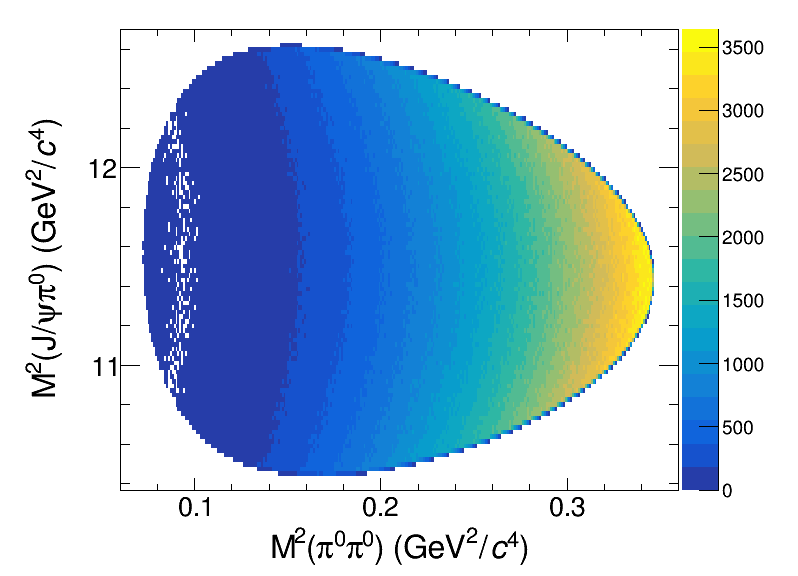}
    \caption{Dalitz plot of $M^{2}(\pi^{0}\pi^{0})$ versus $M^{2}(J/\psi\pi^{0})$ for the data of the decay $\psi(3686) \to J/\psi \pi^{0} \pi^{0}$.}
    \label{fig:dalitz}
\end{figure}

To ensure that this threshold enhancement does not arise from background contributions, extensive studies are performed using both Monte Carlo (MC) simulations and data. 
Possible background contributions are investigated using an inclusive $\psi(3686)$ MC sample corresponding to 2.7 billion \(\psi(3686)\) decays, generated with {\sc lundcharm}~\cite{Chen:2000tv, Yang:2014vra} and {\sc evtgen}~\cite{Ping:2008zz, Lange:2001uf}, together with the initial-state radiation (ISR) production of the $J/\psi$ and continuum processes incorporated in {\sc kkmc}~\cite{Jadach:2000ir}. The background contamination rate is estimated to be about 0.1\%, mainly coming from $\psi(3686)\rightarrow\eta J/\psi$ with $\eta\rightarrow 3\pi^0$. The contribution from non-resonant $e^+e^-$ annihilation is estimated using a data sample collected at $\sqrt{s}=3.65~{\rm GeV}$ with an integrated luminosity of 445.5~$\rm{pb}^{-1}$. None of the background sources produces a peaking structure near the $\pi^0\pi^0$ mass threshold, as illustrated in Fig.~\ref{fig:bkgpipi}. In the Dalitz plot, a clear threshold enhancement in the $M(\pi^0\pi^0)$ distribution is observed, as shown in Fig.~\ref{fig:dalitz}.

\section{Fit to the mass spectrum}

\begin{figure}[htbp]
    \centering
    \includegraphics[width=0.42\textwidth]{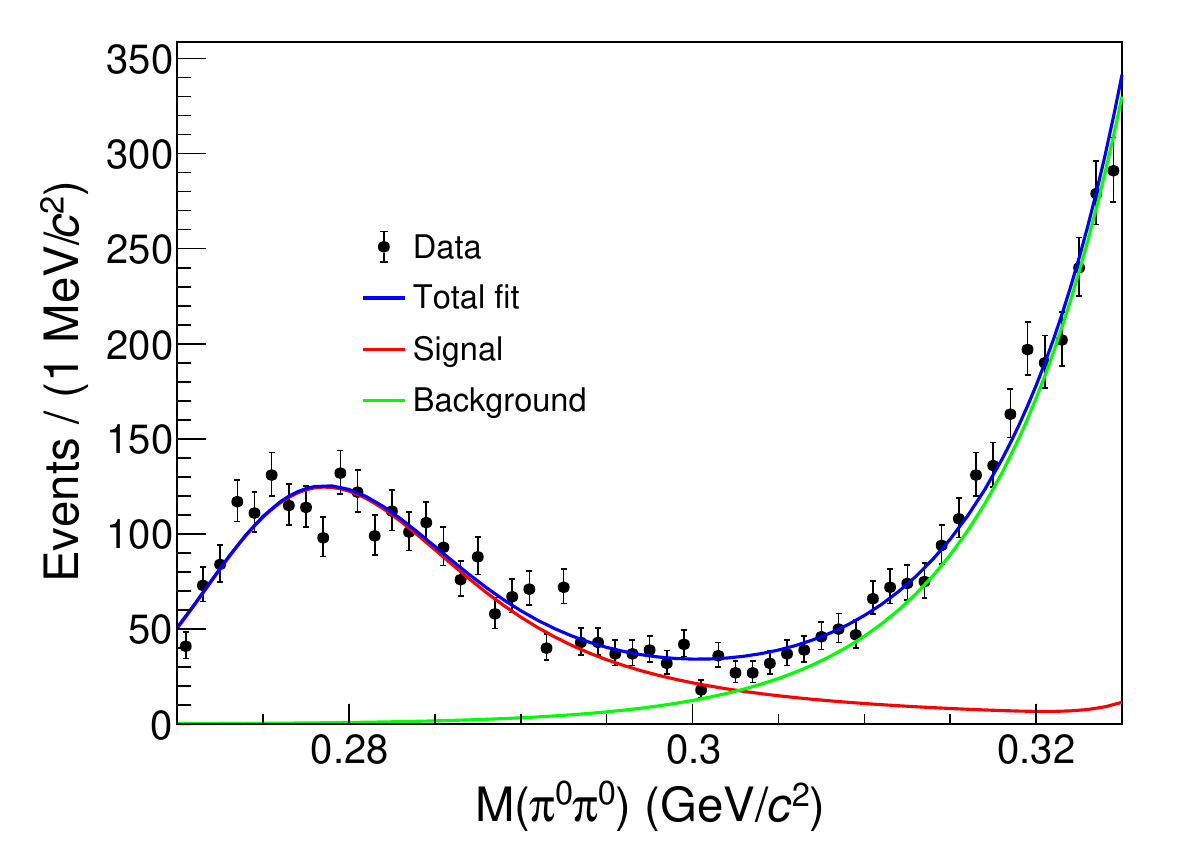}
    \caption{The fit to the $M(\pi^0\pi^0)$ distribution. The dots with error bars represent data, the blue line represents the total fit, the red line represents the signal, and the green line represents the non-peaking contribution from genuine $\psi(3686)\to \pi^0\pi^0 J/\psi$ decays outside the threshold region.
    }
    \label{fig:pipifit}
\end{figure}

An unbinned maximum-likelihood fit to the $\pi^0\pi^0$ mass spectrum is performed. The signal is modeled by a Breit-Wigner function, multiplied by the two-body phase-space factor and the detection efficiency, and convolved with a Gaussian resolution function. The contribution from the left tail of the broad structure above 0.3 GeV/$c^2$ is modeled by an exponential function. As shown in Fig.~\ref{fig:pipifit}, the fit yields $2554\pm62$ signal events, with a mass of $274.3\pm 2.1~{\rm MeV}/c^2$ and a width of $17.1\pm 1.5~{\rm MeV}$, where the uncertainties are statistical only. The fitted mass and width are consistent with those obtained from $\psi(3686)\rightarrow\pi^+\pi^- J/\psi$~\cite{BESIII:2025ozb}.

\section{Theoretical analysis}

To enable direct comparison with theoretical models, the background-subtracted binned spectrum is unfolded using a response matrix derived from signal MC simulation. The unfolding employs the iterative Bayesian method~\cite{DAgostini:1994fjx, Richardson:1972hli, Lucy:1974yx}, as implemented in the RooUnfold software package~\cite{Brenner:2019lmf, Adye:2011gm}, with four iterations. The unfolded spectrum is subsequently used in a two-dimensional binned maximum-likelihood fit to extract the model parameters.
Data from both the $J/\psi\rightarrow e^+e^-$ and $J/\psi\rightarrow\mu^+\mu^-$ decay channels are fitted simultaneously.
  
The fit is performed using the $\pi^0\pi^0$ invariant mass and the $\pi^0$ helicity angle, $\cos\theta_{\pi^0}$, as the fit variables. Here, $\cos\theta_{\pi^0}$ is defined as the angle between the momentum of the $\pi^{0}$ in the rest frame of the $\pi^{0}\pi^{0}$ system and the momentum of the $\pi^{0}\pi^{0}$ system in the rest frame of the initial $\psi(3686)$.
The unfolded spectrum is interpreted within two theoretical frameworks: ChPT~\cite{Mannel:1995jt, Guo:2004dt, Chen:2015jgl} and QCDME~\cite{Kuang:2006me,Kuang:2012wp}.

Within the framework of ChPT~\cite{Mannel:1995jt, Guo:2004dt, Chen:2015jgl}, the differential decay rate is
\begin{equation}
    \frac{d\Gamma}{dm_{\pi\pi}~d\cos{\theta_{\pi^0}}}=\frac{m_{\pi\pi}\sigma_{\pi}|\bf{q}|}
    {64\pi^{3}m_{\psi}^{2}}|\mathcal{M}_S+\mathcal{M}_D|^{2},
\end{equation}
where $|{\bf q}|=\frac{1}{2m_{\psi}}\lambda^{1/2}(m_{\psi}^{2}, m_{J/\psi}^{2}, m^{2}_{\pi\pi})$ is the magnitude of the three-momentum of the $\pi\pi$ system in the $\psi(3686)$ rest frame, with $\lambda(a,b,c)=a^2+b^2+c^2-2ab-2ac-2bc$ and $\sigma_{\pi}=\sqrt{1-4m_{\pi}^{2}/m_{\pi\pi}^2}$.
Here, $\mathcal{M}_S$ and $\mathcal{M}_D$ are the $S$-wave and $D$-wave amplitudes, respectively~\cite{BESIII:2025ozb}. The couplings $g_{0}$ and $g_{1}$ ($g_{2}$) are the free parameters of the $S$-wave ($D$-wave) amplitude, whose explicit expressions are those of
Eq.~(8) of Ref.~\cite{Mannel:1995jt}. Since only the overall normalisation is observable, only the ratios $g_{1}/g_{0}$ and $g_{2}/g_{0}$ are determined by the fit.

\begin{figure}[h]
    \centering
    \subfigure{
        \begin{overpic}[page=1, width=0.45\textwidth]{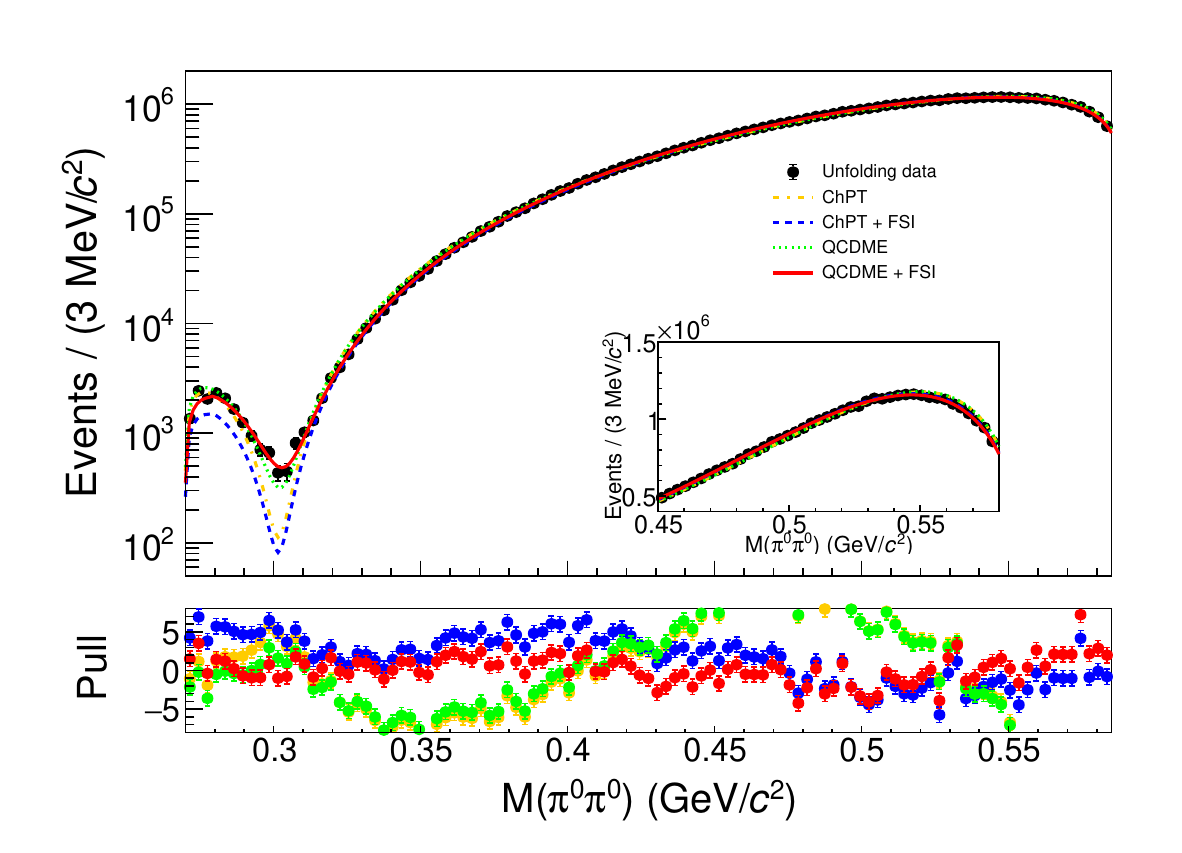}
        \put(18,62){\bf (a)}
        \end{overpic}
    }
    \subfigure{
        \begin{overpic}[page=2, width=0.45\textwidth]{compare.pdf}
        \put(18,62){\bf (b)}
        \end{overpic}
    }
    \caption{The fitting results of (a) $M(\pi^0\pi^0)$ and (b) $\cos{\theta_{\pi^0}}$ with four theories. The dots are data, the yellow dash-dotted curve is the ChPT theory fitting results, the blue dotted curve is the ChPT theory with FSI fitting results, the green dashed curve is the QCDME fitting result and the red curve is the QCDME with FSI fitting result. The pull distribution is derived from the fit results of the theoretical models and the data.}
    \label{fig:totalfit}
\end{figure}

As shown in Fig.~\ref{fig:totalfit}, a simultaneous fit to the $\pi^0\pi^0$ mass spectrum and the $\cos\theta$ angular distribution yields significant discrepancies in both the low- and high-mass regions.
Incorporating FSI within the chiral unitary approach (CHUA)~\cite{Guo:2004dt, Liu:2012dv, Oller:2000ma}, the leading-order $S$-wave amplitude is corrected to $\mathcal{M}_S'=\mathcal{M}_S\cdot(1+  \frac{1}{16\pi^{2}}[ \gamma+\sigma_{\pi}\ln{ \frac{\sigma_{\pi}-1}{\sigma_{\pi}+1} } ] \cdot t_{\pi\pi})$, where $t_{\pi\pi} = (m_{\pi\pi}^2-m_\pi^2)/F_{\pi}$ and $F_\pi$ is the pion decay constant. Here $\gamma$ is a subtraction constant that regularizes the loop. 
The data is then refitted using the modified amplitude, yielding $g_{1}/g_0 = -0.09426 \pm 0.00002$, $g_{2}/g_0 = (0 \pm 7)\times 10^{-2}$, and $\gamma = -3.713 \pm 0.001$. 
Based on the fitted projection of the $\pi^0\pi^0$ mass spectrum as illustrated in Fig.~\ref{fig:totalfit}, the consistency between data and fit is significantly improved in the high-mass region around 0.55~GeV/$c^2$, with $\chi^2$/ndf reduced from 48.6 to 6.5. 

In accordance with QCDME, the $\psi(3686)$ is modeled as an admixture of $\psi(2S)$ and $\psi(1D)$ states. The differential decay width can be described, following Refs.~\cite{Kuang:2012wp, Yan:1980uh}, as
\begin{equation}
\begin{split}
    \frac{d\Gamma}{dm_{\pi\pi}d\cos{\theta_{\pi^0}}} =& \cos^{2}{\theta_{\rm mix}}
    \frac{dG_{\alpha}}{d m_{\pi\pi}d\cos{\theta_{\pi^0}}} \\
    &+ \sin^{2}{\theta_{\rm mix}} \frac{dH_{\beta}}{d m_{\pi\pi}d\cos{\theta_{\pi^0}}},
\end{split}
\end{equation}
where $\theta_{\rm mix}$ is the mixing angle between $\psi(3686)$ and $\psi(3770)$. Similarly, $\alpha$ and $\beta$ are the couplings of the $S$-wave $G_{\alpha}$ and $D$-wave $H_{\beta}$ amplitudes, respectively, whose explicit expressions are those of Eq.~(34) of Ref.~\cite{Kuang:2012wp}.

In this case, we observe a strong correlation between the mixing angle $\theta_{\rm mix}$ and the parameter $\beta$.
Fixing the mixing angle to $\theta_{\rm mix}=12^{\circ}$, as determined from the ratio of electronic widths~\cite{Kuang:2006me}, the fit (shown in Fig.~\ref{fig:totalfit}) reproduces the threshold enhancement but overshoots the spectrum near 0.55 GeV/$c^2$, reducing $\chi^2$/ndf from 4.8 to 3.4. 

To account for the FSI contribution, the same FSI amplitude as described above is applied to $G_\alpha$. The resulting fit yields 
$\alpha = -0.09516 \pm 0.0005_{\rm stat.} \pm 0.0039_{\rm syst.}$, 
$\beta = 0.368 \pm 0.004_{\rm stat.} \pm 0.007_{\rm syst.}$, 
and $\gamma = -3.556 \pm 0.004_{\rm stat.} \pm 0.032_{\rm syst.}$.
As shown in Fig.~\ref{fig:totalfit}, the fitted projection is in agreement with the data, implying that QCDME combined with FSI provides a significantly improved description of the data. The $\cos\theta_{\pi^0}$ angular projection [Fig.~\ref{fig:totalfit}(b)] also shows better agreement than the ChPT model, though a clear systematic trend remains in the $\cos\theta_{\pi^0}$-dependent pull distribution, suggesting a modest limitation of the leading-order angular parameterization.

\section{Systematic uncertainties}

Sources of systematic uncertainty in the fitting parameters are summarized in Table~\ref{tab:casec}. The uncertainties arise from data-MC differences (MDC tracking, photon detection, kinematic fit, and resolution), as well as from the requirements on $\cos{\theta}$. The data-MC efficiency difference in MDC tracking is studied using control samples of $\psi(3686)\rightarrow \pi^+\pi^-J/\psi$ with $J/\psi\rightarrow e^+e^-$ or $J/\psi\rightarrow\mu^+\mu^-$. The difference for EMC photon detection is studied using control samples with $e^+e^-\rightarrow \gamma_{\rm ISR} \mu^+\mu^- $. The difference in fit results between efficiencies with and without correction is taken as the systematic uncertainty. In accordance with the previous BESIII publication~\cite{BESIII:2012mpj}, the efficiency with the track-helix correction is taken as the nominal value, and the difference in fit results between efficiencies with and without this correction is assigned as the systematic uncertainty from the kinematic fit. The effect of the mass constraints is probed by repeating the selection and the fit with the 4C kinematic fit, in which the $J/\psi$ and the two $\pi^0$ mass constraints are removed; the corresponding change in the fitted parameters is assigned as the mass constraints uncertainty.
The possible photon miscombination in the signal MC sample has been studied by tagging miscombination events through matching the true and reconstructed photon momenta. Two fits are performed to the samples with and without miscombination events, and the change in the results is taken as the systematic uncertainty.
For the requirements on $\cos{\theta}$ of leptons, the discrepancies in the fit parameters with and without this selection criterion are regarded as systematic uncertainties.
For the resolution, the MC resolution is increased by 10\%, and the changes in the fit results are taken as systematic uncertainties. 
The systematic uncertainty from the binning strategy is found to be negligible.

\begin{table}[htp]
    \caption{The summary of systematic sources and their contributions to the fitted parameters (in percent). The $\alpha$, $\beta$ and $\gamma$ are the fitting parameters of QCDME. The $m_X$ and $\Gamma_X$ are the mass and width with Breit-Wigner function.}
    \centering
    \setlength{\tabcolsep}{1.5mm}{
    \begin{tabular}{c|c c c c c}
    \hline
        Source& $\alpha$ & $\beta$ & $\gamma$ & $m_X$ & $\Gamma_X$ \\
    \hline
        MDC tracking & $\dots$ & 0.5 & 0.2 & $\dots$ & $\dots$\\
        Photon & 0.1 & 0.3 & 0.1 & 0.1 & 0.2 \\
        Kinematic fit & 0.2 & 0.4 & 0.2 & $\dots$ & 0.2\\
        Mass constraints & $\dots$ & 1.4 & 0.8 & $\dots$ & 1.0\\
        Photon miscombination & 0.3 & 0.4 & 0.2  & 0.1 & 0.2\\
        $\cos{\theta}$ requirement & 0.1 & 0.5 & 0.1 & 0.2 & 2.4\\
        Resolution & 0.1 & 0.9 & 0.3 & $\dots$ & 0.2\\
    \hline
        Total & 0.4 & 1.9 & 0.9 & 0.2 & 2.6\\
    \hline
    \end{tabular}}
    \label{tab:casec}
\end{table}

\section{Summary}

In summary, utilizing a data sample of $(2712.4\pm14.4)\times 10^{6}$ $\psi(3686)$ events, we have performed a high-precision study of the $\psi(3686)\rightarrow\pi^{0}\pi^{0}J/\psi$ decay. A clear enhancement near the $\pi^{0}\pi^{0}$ mass threshold is confirmed. A Breit-Wigner fit yields a mass of $M = 274.3\pm 2.1_{\rm stat.}\pm0.6_{\rm syst.}~{\rm MeV}/c^2$ and a width of $\Gamma = 17.1\pm 1.5_{\rm stat.}\pm0.5_{\rm syst.}~{\rm MeV}$, consistent with the values observed in the charged $\pi^{+}\pi^{-}$ channel~\cite{BESIII:2025ozb}. This agreement, established in the clean neutral environment free from initial-state radiation backgrounds, provides independent and robust confirmation of the structure's existence. 

Theoretical analyses using ChPT with FSI and the QCDME model with an $S$- and $D$-wave mixing angle of $12^\circ$ yield conclusions consistent with those from the charged channel: ChPT fails to describe the threshold enhancement, while the QCDME framework successfully reproduces the overall shape of the $\pi^{0}\pi^{0}$ mass spectrum and provides a better description of the $\cos\theta_{\pi^0}$ angular distribution, albeit with a clear systematic trend  $\cos\theta_{\pi^0}$-dependent pull distribution. The persistence of the dip around 0.3 GeV/$c^2$ in the neutral channel reinforces its physical origin, likely linked to the Adler zero requirement. The combined results from both decay channels offer a more complete picture, significantly deepening our understanding of low-energy QCD and the role of chiral symmetry in hadronic transitions of charmonium states.

\section*{Conflict of interest}

The authors declare that they have no conflict of interest.

\section*{Acknowledgments}
The BESIII Collaboration thanks the staff of BEPCII (https://cstr.cn/31109.02.BEPC) and the IHEP computing center for their strong support. This work is supported in part by National Key R\&D Program of China under Contracts Nos. 2025YFA1613900, 2023YFA1606000, 2023YFA1606704; National Natural Science Foundation of China (NSFC) under Contracts Nos. 11635010, 11935015, 11935016, 11935018, 12025502, 12035009, 12035013, 12061131003, 12192260, 12192261, 12192262, 12192263, 12192264, 12192265, 12221005, 12225509, 12235017, 12342502, 12361141819, 12535005; the Chinese Academy of Sciences (CAS) Large-Scale Scientific Facility Program; the Strategic Priority Research Program of Chinese Academy of Sciences under Contract No. XDA0480600; CAS under Contract No. YSBR-101; 100 Talents Program of CAS; The Institute of Nuclear and Particle Physics (INPAC) and Shanghai Key Laboratory for Particle Physics and Cosmology; Agencia Nacional de Investigación y Desarrollo de Chile (ANID), Chile under Contract No. ANID CCTVal CIA250027; ERC under Contract No. 758462; German Research Foundation DFG under Contract No. FOR5327; Istituto Nazionale di Fisica Nucleare, Italy; Knut and Alice Wallenberg Foundation under Contracts Nos. 2021.0174, 2021.0299, 2023.0315; Ministry of Development of Turkey under Contract No. DPT2006K-120470; National Research Foundation of Korea under Contract No. RS-2026-25486791; National Science and Technology fund of Mongolia; Polish National Science Centre under Contract No. 2024/53/B/ST2/00975; STFC (United Kingdom); Swedish Research Council under Contract No. 2019.04595; U. S. Department of Energy under Contract No. DE-FG02-05ER41374

\bibliographystyle{elsarticle-num}
\bibliography{main}

\onecolumn
\section*{Author list and affiliations}
\begin{small}
\begin{center}
M.~Ablikim$^{1}$\BESIIIorcid{0000-0002-3935-619X},
M.~N.~Achasov$^{4,c}$\BESIIIorcid{0000-0002-9400-8622},
P.~Adlarson$^{84}$\BESIIIorcid{0000-0001-6280-3851},
X.~C.~Ai$^{90}$\BESIIIorcid{0000-0003-3856-2415},
C.~S.~Akondi$^{32A,32B}$\BESIIIorcid{0000-0001-6303-5217},
R.~Aliberti$^{40}$\BESIIIorcid{0000-0003-3500-4012},
A.~Amoroso$^{83A,83C}$\BESIIIorcid{0000-0002-3095-8610},
Q.~An$^{79,66,\dagger}$,
Y.~H.~An$^{90}$\BESIIIorcid{0009-0008-3419-0849},
M.~S.~Anderson$^{40}$\BESIIIorcid{0009-0008-1550-2632}
% ...
\\
\vspace{0.2cm}
(BESIII Collaboration)\\
\vspace{0.2cm} {\it
$^{1}$ Institute of High Energy Physics, Beijing 100049, People's Republic of China\\
% ...
$^{4}$ Budker Institute of Nuclear Physics SB RAS (BINP), Novosibirsk 630090, Russia\\
% ...
$^{32}$ INFN Sezione di Ferrara, (A)INFN Sezione di Ferrara, I-44122, Ferrara, Italy; (B)University of Ferrara, I-44122, Ferrara, Italy\\
% ...
$^{40}$ Johannes Gutenberg University of Mainz, Johann-Joachim-Becher-Weg 45, D-55099 Mainz, Germany\\
% ...
$^{66}$ State Key Laboratory of Particle Detection and Electronics, Beijing 100049, Hefei 230026, People's Republic of China\\
% ...
$^{79}$ University of Science and Technology of China, Hefei 230026, People's Republic of China\\
% ...
$^{83}$ University of Turin and INFN, (A)University of Turin, I-10125, Turin, Italy; (B)University of Eastern Piedmont, I-15121, Alessandria, Italy; (C)INFN, I-10125, Turin, Italy\\
$^{84}$ Uppsala University, Box 516, SE-75120 Uppsala, Sweden\\
% ...
$^{90}$ Zhengzhou University, Zhengzhou 450001, People's Republic of China\\

\vspace{0.2cm}
$^{\dagger}$ Deceased\\
% ...
$^{c}$ Also at the Novosibirsk State University, Novosibirsk, 630090, Russia\\
% ...

}

\end{center}
\end{small}

\end{document}